# Statistical Study of Spatial Distribution and Polarization for Saturn Narrowband Emissions

Siyuan Wu[1], Shengyi Ye[1], G. Fischer[2], Jian Wang[1], Minyi Long[3], J. D. Menietti[4], B. Cecconi[5], and W. S. Kurth[4]

[1] Department of Earth and Space Sciences, Southern University of Science and Technology, Shenzhen, Guangdong, People's Republic of China; yesy@sustech.edu.cn
[2] Space Research Institute, Austrian Academy of Sciences Graz, Austria
[3] Department of Space Physics, School of Electronic Information, Wuhan University, Wuhan, People's Republic of China
[4] Department of Physics and Astronomy, University of Iowa Iowa City, IA, USA
[5] LESIA, Observatoire de Paris, Université PSL, CNRS, Sorbonne Université, Université de Paris Meudon, France
*Received 2021 April 1; revised 2021 June 7; accepted 2021 June 9; published 2021 MM DD*

## Abstract

The spatial distribution and polarization of Saturn narrowband (NB) emissions have been studied by using Cassini Radio and Plasma Wave Sciences data and goniopolarimetric data obtained through an inversion algorithm with a preset source located at the center of Saturn. From 2004 January 1 to 2017 September 12, NB emissions were selected automatically by a computer program and rechecked manually. The spatial distribution shows a preference for high latitude and intensity peaks in the region within 6 Saturn radii ($R_s$) for both 5 and 20 kHz NB emissions. 5 kHz NB emissions also show a local time preference roughly in the 18:00–22:00 sector. The Enceladus plasma torus makes it difficult for NB emissions to propagate to the low latitude regions outside the plasma torus. The extent of the low latitude regions where 5 and 20 kHz NB emissions were never observed is consistent with the corresponding plasma torus density contour in the meridional plane. 20 kHz NB emissions show a high circular polarization while 5 kHz NB emissions are less circularly polarized with $|V| < 0.6$ for majority of the cases. And cases of 5 kHz NB emissions with high circular polarization are more frequently observed at high latitude especially at the northern and southern edges of the Enceladus plasma torus.

*Unified Astronomy Thesaurus concepts:* Magnetospheric radio emissions (998); Saturn (1426); Planetary magnetospheres (997);

*Supporting material:* machine-readable table

## 1. Introduction

It has been 40 yr since the Voyager 1 Saturn flyby and narrowband (NB) emissions were first observed by the Voyager Plasma Wave System (Scarf & Gurnett 1977) and reported by Gurnett et al. (1981). The locations where Voyager 1 recorded NB emissions varied between 3 and 58 Saturn $R_s$ (60,268 km used in this work), and the emissions were above the plasma frequency ($f_{pe}$). Therefore, it was recognized as the L-O mode. Cassini entered Saturn's orbit on 2004 July 1 and started its 13 yr journey to study the giant gaseous planet. Since then, NB emissions have been frequently observed in the electromagnetic field spectrum measured by the Radio and Plasma Wave Science (RPWS) instrument on board Cassini (Gurnett et al. 2004).

Figure 1(a) gives an example of RPWS measurements with the detected 5 and 20 kHz NB emissions from 2009 May 1–3. NB emissions are also referred to as "n-SMR" emissions (narrowband Saturn myriametric radiation) for signals in the 3–10 kHz frequency range (Louarn et al. 2007), and "n-SKR" in the 10–40 kHz range (Lamy et al. 2008) and more recently 5 and 20 kHz NB emissions (Ye et al. 2009; Wang et al. 2010; Menietti et al. 2019; Wing et al. 2020).

Unlike Saturn Kilometric Radiation (SKR), which is directly generated via the cyclotron maser instability (CMI; Wu & Lee 1979; Lamy et al. 2008), NB emissions are believed to be generated through mode conversion. Kurth et al. (1981) showed the evidence that Earth's continuum radiation can be mode converted from upper hybrid waves. The generation mechanism of NB emissions was first proposed by Gurnett et al. (1981) to be mode conversion from electrostatic waves near the upper hybrid resonance via either a linear (Jones 1976) or nonlinear mechanism (Melrose 1981) based on its similarities with terrestrial narrow-banded nonthermal continuum emissions. Ye et al. (2009) reported such observations when Cassini crossed the source region of 20 kHz NB emissions, which is indicated by a radio emission preceding or following a strong electrostatic wave. The modeled source locations based on the mode conversion theory of Jones (1976) and Melrose (1981) showed a very good agreement with the observed 20 kHz NB emission source regions. However, the 5 kHz NB emissions source locations obtained through radio wave direction finding (DF; Cecconi & Zarka 2005) did not agree with the model (Ye et al. 2009), which implies a different source mechanism.

Farrell et al. (2005) first reported Z mode NB emissions observed interior to the plasma torus and discussed the possible mode conversion generation of the Z mode from Langmuir waves through the whistler branch. Menietti et al. (2009) gave an example of 20 kHz Z mode NB emissions near a possible source region, and by doing a linear growth rate calculation, suggested Z mode NB emissions could be directly generated by a CMI mechanism when the condition $f_{uh} = nf_{ce}$, ($f_{uh}$: upper hybrid frequency, $f_{ce}$: electron cyclotron frequency) is met. The

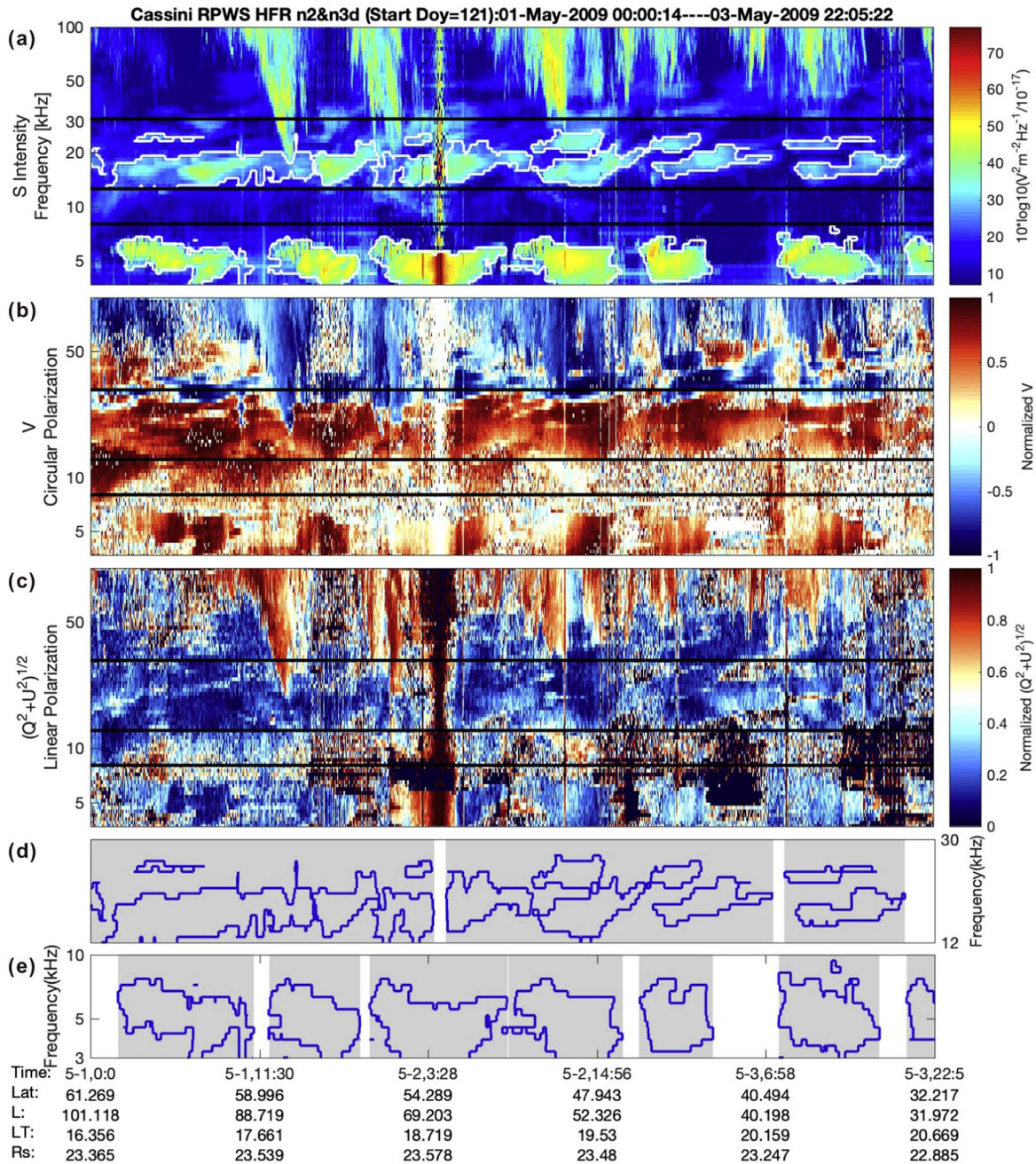

**Figure 1.** Panel (a) wave electric field spectrogram measured by the Cassini RPWS High Frequency Receiver (HFR) in the frequency range from 3 – 100 kHz. The horizontal lines represent the frequencies at 8, 12, and 30 kHz; and the white contour lines correspond to the selected 5 and 20 kHz NB emission cases. Panels (b) and (c) give the circular polarization and linear polarization degrees with 1 representing left-hand circular (LHC) polarization, −1 representing right-hand circular (RHC) polarization, and 1 in panel (c) for full linear polarization, 0 for no linear polarization. Panels (d) and (e) give detailed features of the results for 20 and 5 kHz NB emissions automatically selected by our computer program. The contour lines are the same as that in panel (a). The gray areas mark the contour lines for clarity. The horizontal ephemeris data give the universal time (Time), latitude (Lat), L shell values (L), local time (LT), and radial distance ($R_s$) in units of Saturn radii.

Z mode can mode convert to L-O mode at density gradients and finally be observed by Cassini (Ye et al. 2009; Menietti et al. 2018b).

Ye et al. (2010b) discovered 5 kHz Z mode NB emissions and proposed the mode conversion mechanism from the Z mode to the L-O mode. Since the Z mode can propagate between the isosurfaces of $f_{uh}$ and $f_{L=0}$ (lower cutoff for the Z mode) frequencies, this work also gave a model of the region where the Z mode can propagate, which matched the observed Z mode distribution well (Ye et al. 2010b; Menietti et al. 2015). Statistical studies of 5 kHz Z mode emissions show that these Z mode emissions are distributed in the low density region inside the inner edge of the Enceladus plasma torus in the lower latitude region within 3 $R_s$ and extend to approximately 7 $R_s$ above and below the equatorial plane (Menietti et al. 2018a, 2018b). The intensity of the Z mode peaks near 25° latitude with radial distance very close to Saturn, and they are much more intense relative to O mode NB emissions at larger radial distances (Menietti et al. 2015). The stochastic modeling of electron scattering indicates that the Z mode at Saturn is dominantly responsible for the electron acceleration to megaelectronvolt energies inside 4 $R_s$, i.e., responsible for the formation of the radiation belt inside the Enceladus' orbit (Gu et al. 2013; Woodfield et al. 2018).

Recently, Menietti et al. (2016, 2019) studied the generation mechanism of Z mode emissions for both 5 and 20 kHz and the relation of upper hybrid resonance with NB emissions by analyzing the growth rates of the waves based on the electron phase space distribution measured by the Cassini Plasma Spectrometer (CAPS) Electron Spectrometer (ELS; Young et al. 2004). Case studies show the wave normal angel of the Z mode to be usually smaller than 30° (Menietti et al. 2018a, 2018b), and DF results show two possible source regions: one at high latitude on the same magnetic field lines as SKR and another one at the equatorial region interior to the plasma torus (Menietti et al. 2018b). It was found that the loss-cone and temperature anisotropies in the electron distribution are responsible for the excitation of the Z modes and other concomitant waves (Menietti et al. 2016). In their work, they also found that the Z mode and O mode can be directly generated in the SKR source region (Menietti et al. 2011). Wing et al. (2020) suggested that 5 kHz NB emissions could be related to the energetic neutral atom (ENA) brightening events, which have been interpreted as plasma injections in the Kronian magnetosphere.

NB emissions have been extensively studied by using earlier Cassini RPWS data, including their spatial distribution (Ye et al. 2009), intensity (Wang et al. 2010), polarization (Ye et al. 2010b), source location (Ye et al. 2009), and excitation mechanism (Menietti et al. 2009, 2019; Ye et al. 2009). However, earlier studies were based on data obtained from a limited orbit coverage and many studies used only cases to study other characteristics such as the generation mechanism (Menietti et al. 2016), source location (Ye et al. 2009), and wave particle interaction (Gu et al. 2013; Woodfield et al. 2018). The spatial distribution and polarization features of NB emissions were analyzed based on partial data by Wang et al. (2010) and Ye et al. (2009, 2010b), and for the Z mode by Menietti et al. (2018a, 2018b). The location where Cassini observed the NB emissions and the latitude-local time occurrence rates profile given by Wang et al. (2010) were based on only 2 yr of Cassini data from 2005 September to 2007 May. The spatial coverage of Cassini at that time was very limited, with only dayside local times in the southern hemisphere and nightside local times at the northern hemisphere.

The polarization of NB emissions has not yet been systematically analyzed. It has been reported (Ye et al. 2009) that NB emissions are purely circularly polarized at high latitudes but partially polarized or unpolarized at low latitude with no linear component (Fischer et al., presented paper, 2008). However, these results have not been published formally. Some researchers use the goniopolarimetric (DF) data to identify the wave mode case by case since NB emissions are believed to be the L-O mode and Z mode, which would have a circular polarization with different senses (Ye et al. 2010b; Woodfield et al. 2018; Menietti et al. 2019). No full analysis of the 5 and 20 kHz NB emissions' polarization has been published. Fischer et al. (2009) found that the SKR is elliptically polarized at high latitude and stressed the carefulness of using the goniopolarimetric data when the inversion assumes a purely circularly polarized wave. Therefore, it is important to carry out a statistical study on this topic. The details of the goniopolarimetric parameters will be introduced in Sections 2 and 3.

Above all, the source location of NB emissions is still not clear, the spatial distribution and polarization parameters still need a comprehensive understanding. In this study, we present a statistical study of 5 and 20 kHz NB emissions for the spatial distribution and polarization characteristics by using the Cassini RPWS data from 2004−2017. The data and criteria adopted to identify the NB emission events are described in Section 2. The results are presented in Sections 3 and 4, and finally the discussion and conclusion are in Section 5.

## 2. Data and Method

This work used the data obtained by the Cassini RPWS High Frequency Receiver (HFR), which recorded the electric field over a frequency range of 3.5 kHz–16 MHz (Gurnett et al. 2004). The data used in this work covers 2004, day 001 to 2017, day 255 (day of year (DOY)). The data coverage is almost full with only several days missing. The magnetic field data is from the Cassini MAG (magnetometer) instrument with 1 minute resolution (Dougherty et al. 2004). The Stokes parameters are obtained from the DF method using two antennas with a preset source location at the center of Saturn (Cecconi & Zarka 2005). All spectrogram data are averaged over 1 minute intervals.

5 kHz NB emission events have a maximum intensity near 5 kHz (Menietti et al. 2009), and early studies used the frequency range over 3–8 kHz to identify the NB emission cases (Ye et al. 2010a, 2010b). The measured electric field intensities are averaged over a frequency range between 4.4 and 5.6 kHz in this work. Therefore, when the NB emissions are detected by the RPWS instrument, the frequency averaged time series would show an intensity enhancement. Compared to the intensity at edge frequency channels (3 and 8 kHz in this work), the intensity at the center frequency channels should be stronger than that at edge channels when NB emission signals are received. Therefore, we use a parameter $P$ to locate the NB emission event. Whenever the parameter $P$ gets larger than a threshold value, the data would be tagged as an NB emission event. Parameter $P$ is defined as

$$P = \frac{\text{mean}(\text{Intensity}_{[4.4\,\text{kHz} \sim 5.6\,\text{kHz}]})}{\text{minimum}(\text{Intensity}_{3\,\text{kHz},8\,\text{kHz}})}.$$

The denominator is defined as the minimum value for each time step at either edge frequency channels (3 or 8 kHz). One can easily understand that this $P$ value could be used to represent the intensity enhancement for a single measurement time at the central frequency channels relative to the intensity at the edge frequency. This research uses a threshold value of 1.15 as the intensity enhancement level of a NB emission event. We tested various threshold levels and found that $P > 1.15$ yielded a good time range of 5 kHz NB emissions in the dynamic spectra. We note here that this filter process by using parameter $P$ is not a peak detection algorithm since this process will pick every measurement with an intensity enhancement no matter the intensity of the signal itself is strong or weak.

Once the time ranges of 5 kHz NB emissions were identified, we next found the areas with intensities larger than 23 dB above the background (the background is set to be $10^{-17}$ V$^2$ m$^{-2}$ Hz$^{-1}$ for 3–30 kHz) at each NB emission event time range on the dynamic spectrum and get the contour lines in Figures 1(a) and (e). This step was operated over a frequency range of 3–8 kHz. Subsequently, the intensity, polarization, and ephemeris data would be saved within the NB emission case interval by averaging the data inside the contour lines for every minute. Finally, the results would be rechecked visually by using dynamic spectra. Examples of the identification are shown in Figures 1(a), (d), and (e). The white and blue contour lines give the areas of identified 5 and 20 kHz NB emissions.

In contrast to the 5 kHz NB emission case, the morphology and frequency range of 20 kHz NB emissions are quite complicated with a diffuse boundary and weak intensity (Wang et al. 2010; Menietti et al. 2018a). The features of 20 kHz NB emissions change from case to case and are often interfered by SKR. We used the criteria below to pick out the 20 kHz NB emissions from the spectrogram within the frequency range of 12–30 kHz: (1) eliminate the data with an intensity below 23 dB, (2) eliminate the data with circular polarization degree smaller than 0.3, (3) intersect the time-frequency area with both intensity enhancement and large circular polarization coefficient, (4) eliminate the data in the northern hemisphere with RHC polarization and LHC polarization in the southern hemisphere at latitudes larger than 10°, (5) eliminate the chosen areas with durations shorter than 1 hr, and (6) manually remove the wrong cases (e.g., the electron cyclotron harmonics and other interferences).

20 kHz NB emissions usually carry a high circular polarization degree, and with an opposite polarization sense to SKR, as NB emissions are L-O mode emissions compared to R-X mode SKR. We treated the 20 kHz NB emissions as circularly polarized with a circular polarization larger than 0.3 and this corresponds to criterion (2). After the filter process we rechecked the results visually and confirmed the validity of this criterion. The SKR would be circular polarized with $V \approx -1$ (right hand) in the northern hemisphere and $V \approx 1$ (left hand) in the southern hemisphere since the angle between wavevector ($k$) and magnetic field ($B$) exceeds 90° in the southern hemisphere (Lamy et al. 2008; Ye et al. 2009). Therefore, criterion (4) is adopted to avoid the contamination with SKR and right-hand polarized second harmonics of the 20 kHz NB emissions to get a more accurate result. In contrast to 20 kHz NB emissions, most of the 5 kHz NB emissions are unpolarized so we do not limit the circular polarization on 5 kHz NB emissions. We note that NB emissions can also be observed above our frequency limit of 30 kHz, and that at such frequencies some NB emissions can also have the same polarization sense as SKR (Ye et al. 2011). As illustrated in Figure 1(d), the contour lines show the 20 kHz NB emission areas on the spectrogram chosen by using the criteria described above. The blue contour lines in panel (d) are the same as the white contour lines in panel (a) of Figure 1.

However, 20 kHz Z mode emissions have usually a very small polarization degree. Parts of the Z mode with circular polarization larger than 0.3 were selected through the criteria above. Therefore, we eliminated all the Z mode emissions chosen by the program and reprocessed all the data to recheck the Z mode emissions by using the method below. The Z mode emissions are usually observed close to Saturn and near the inner boundary of the plasma torus. At low latitudes and close to the plasma torus, the plasma might be in the overdense ($f_{pe} > f_{ec}$) regime, the L-O mode emissions will cutoff at $f_{pe}$ which is larger than the $f_{ec}$. So the Z mode emissions in these regions below $f_{ec}$ will be identified. When the Z mode emissions are observed at high latitude, the plasma is in the underdense ($f_{pe} < f_{ec}$) case, Z mode emissions below local $f_{ec}$ will show an RHC polarization relative to the L-O mode. We changed the algorithm by removing criteria (2) to (4) and set the intensity threshold to 40 dB since Z mode emissions usually show much higher intensity. Then the Z mode emissions are chosen if a case is observed below $f_{ec}$. And the wave mode is further confirmed in combination with the information from the circular polarization.

This work deals with the data for a total observation time start from 2004, 001 to 2017, 255. Finally 2076 (L-O mode) 5 kHz NB emission cases and 1136 (L-O mode) 20 kHz NB emission cases are obtained. We also identified 90 Z mode emissions and 23 Z mode emissions for 5 and 20 kHz NB emissions, respectively. The total duration of NB emission events identified through the program are approximately 8548.8 hr (≈356.2 Earth days) and 5743.2 hr (≈239.3 Earth days) for 5 and 20 kHz NB emissions, respectively. The total observation time is 118,035.3 hr (≈4918 Earth days), yielding total occurrence rates of 7.2% and 4.9% for 5 and 20 kHz NB emissions, respectively.

### 3. Occurrence Rates and Intensity of NB Emission Events

This work gives a statistical study of the spatial distribution of 5 and 20 kHz NB emissions. It has been noted earlier that NB emission events are frequently observed at high latitudes (Wang et al. 2010). However, no full view of the distribution features has been obtained.

Figure 2 shows the distributions of both 5 and 20 kHz NB emission events in the meridional and equatorial planes with each bin representing a 2 $R_s$*2 $R_s$ square. The occurrence rates are calculated within each bin in each coordinate direction by using

$$\text{occurrence rates} = \frac{\text{valid NB points(minutes)}}{\text{observation time(minutes)}}.$$

The valid NB emission points are identical to the data time length of NB emissions that have been observed by Cassini in the corresponding bin (one data point each minute). The observation time is the total time Cassini stayed in that bin. We eliminated the bins with an observation time smaller than 1 day. The detailed parameters for calculating the occurrence rates including observation time and the valid NB points are in the Appendix (Figure A1). The high latitude preference for both 5 and 20 kHz NB emissions are obvious in Figures 2(a) and (c). Since the L-O mode cuts off at $f_{pe}$ which is a function of the plasma density, no NB emission event is observed inside the Enceladus plasma torus (Persoon et al. 2006), i.e., the gap in Figures 2 and 3 panels (a) and (c) at latitude <10° and radial distance between 3 and 15 $R_s$. The region in the plasma torus where $f_{pe}$ is larger than 5 kHz is reflected by the occurrence boundary of the 5 kHz NB emissions, and the region where $f_{pe}$ is larger than 20 kHz is reflected by the occurrence boundary of the 20 kHz NB emissions. The 20 kHz NB emissions can propagate deeper into the plasma torus, which can be seen in Figures 2(a) and (c) in which the 20 kHz NB emissions have a smaller plasma torus envelope than the 5 kHz NB emissions. A more detailed view will be shown later in the discussion in Section 5.

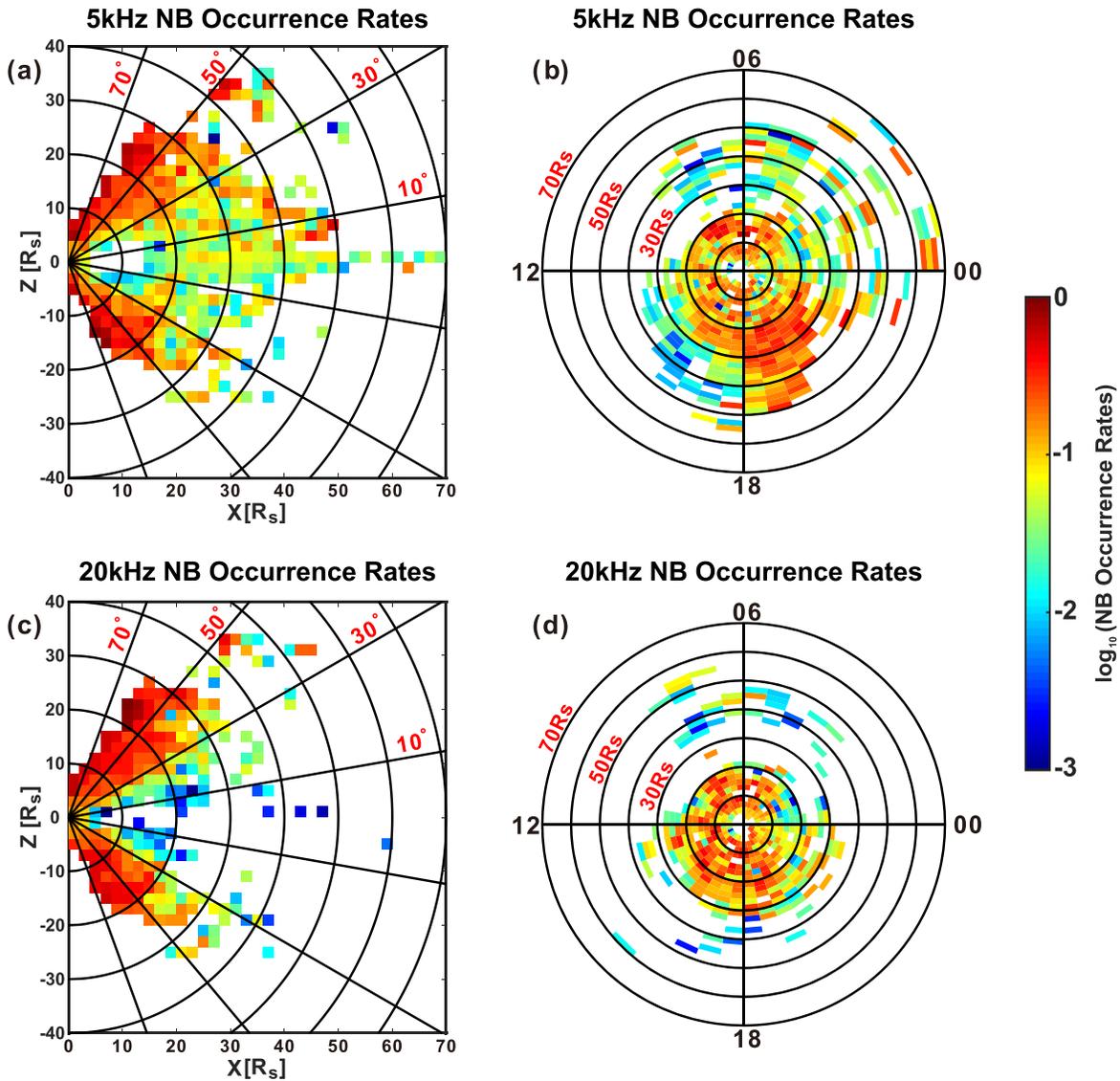

**Figure 2.** Panels (a) and (b) give the 5 kHz NB emission occurrence rates in the meridional and equatorial planes and similarly in panels (c) and (d) for 20 kHz NB emissions. The solid black circles in all panels mark the distance to Saturn from $10-70$ $R_s$. The oblique line in panels (a) and (c) mark the latitudes for $\pm 10°$, $\pm 30°$, $\pm 50°$, and $\pm 70°$.

The 5 kHz NB emissions in Figure 2(b) show a local time preference in the evening sector at roughly $18:00 \sim 22:00$ LT, which is consistent with earlier results (Wang et al. 2010). The 20 kHz NB emissions show a slightly local time preference for the afternoon and evening side compared to the early and late morning side, and this would be a subject of future study.

In Figure 2(d), there is a gap approximately for radial distances from $20-35$ $R_s$ and for local time $1:00-12:00$. And there is a similar pattern also in panel (b) for 5 kHz NB emissions at roughly the 6:00–11:00 local time sector. These gaps are not due to the orbit coverage as one can see in the Appendix. The reason why this gap exists is not clear and needs further study.

Figure 3 shows the distribution of NB emissions' averaged intensity in the equatorial plane and meridional plane. The intensity is averaged within each bin by using all filtered NB emission points, and we note that the intensity is not distance normalized. The most prominent feature is the intensity peak near Saturn within 6 $R_s$ for both 5 and 20 kHz NB emissions. To show the full picture of the NB emissions distribution, we plotted both modes (L-O and Z) in Figures 2–4. The strongest intensity within 6 $R_s$ is mainly due to the NB Z mode emissions, and a closeup of these regions will be given in Section 5. The intensity decreases gradually as the distance to Saturn increases. The intensity outside the plasma torus and at a rather low latitude below $30°$ is weaker than for the NB emissions observed at high latitudes above $30°$. In the equatorial projections in Figures 3(b) and (d), the pattern of decreasing intensity with increasing distance increases is clear. This work also gives a similar result (see Figures A2 and A3 in the Appendix) for the distance-power distribution of the NB emission signals to Figure 2 of Ye et al. (2010b). The results (in Figure A2) support the $1/R^2$ dependence of the wave power to the radial distance as described in their work. However, Figure A3 shows that the 5 kHz NB emissions at low latitudes (within $\pm 10°$) and exterior to the Enceladus plasma torus do not have a clear $1/R^2$ dependence and rather appear to be constant.

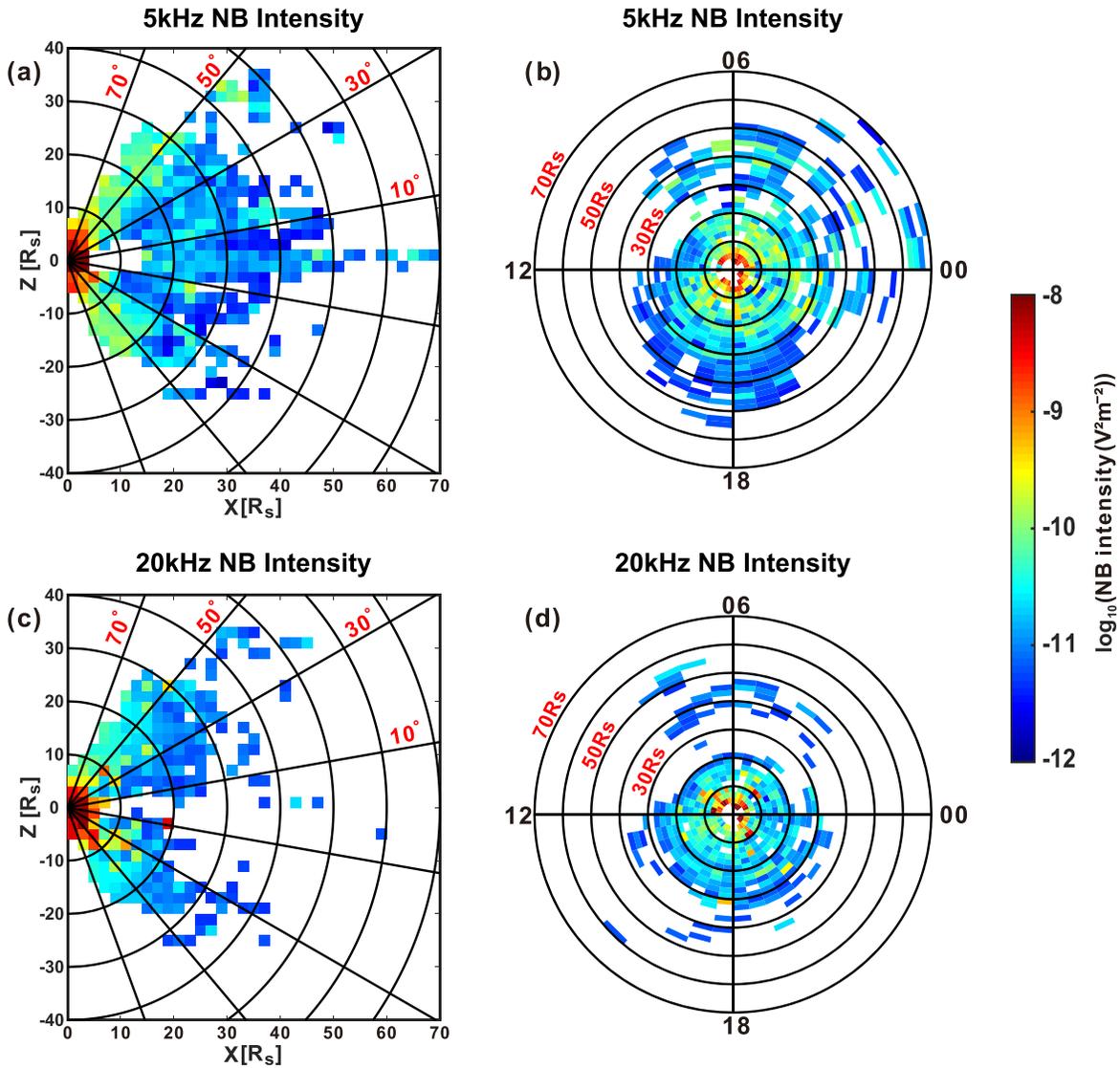

**Figure 3.** Similar format compared to Figure 2 with the color representing averaged intensity.

## 4. Polarization of NB Emission Events

When considering the polarization of waves in a cold plasma, the polarization state of a wave mode is usually defined in the plane perpendicular to the background magnetic field. Therefore, the parallel propagation wave mode such as the L mode and R mode would show a fully LHC and RHC polarization. For perpendicular propagation, the O mode and X mode would be recognized as linear polarized along the ambient magnetic field direction (Cecconi 2019). However, in the plane perpendicular to the wavevector, the polarization of the L-O mode, R-X mode, and Z mode are all circular polarized. The Cassini RPWS instrument has the capability to retrieve Stokes parameters and incoming wave directions from its measurement, which is usually called goniopolarimetry or DF in other studies (Cecconi & Zarka 2005; Cecconi et al. 2009). The details of wave Stokes parameters can be found in Kraus (1966) and the theory basically involves six parameters: intensity ($S$), linear polarization ($Q$ and $U$), circular polarization ($V$), and source direction (angles $\theta$ and $\phi$). The HFR of RPWS usually works in the dipole mode and outputs four values (two autocorrelation and the real part and imaginary parts of the cross correlation for the X antenna and Z antenna) (Gurnett et al. 2004). To calculate the six parameters (Stokes parameters, $\theta$ and $\phi$) by using four measurements, the algorithm needs to make assumptions on two parameters to get the others. Usual approaches are either to preset the source location ($\theta$ and $\phi$ at the direction of the center of Saturn) or to assume the polarization of the wave is purely circular ($U = 0$ and $Q = 0$). The inversion used in this work is also called polarimeter inversion and uses the center of Saturn as a preset source location (Ye et al. 2010b). Figure 4 gives the circular polarization proportion for the NB emissions, the proportion in each bin is calculated through

$$\frac{\text{valid NB emission points(minutes) with } |V| \geqslant 0.6}{\text{valid NB emission points(minutes)}}.$$

Figures 4(a)–(c) show the circular polarization (Stokes V), linear polarization ($\sqrt{Q^2 + U^2}$), and total polarization ($\sqrt{Q^2 + U^2 + V^2}$) proportion of 5 kHz NB. Since we have almost 356.2 Earth days data points (in minutes) of 5 kHz NB emission data and approximately 239.3 Earth days data points of 20 kHz NB emissions (in minutes), instead of showing the

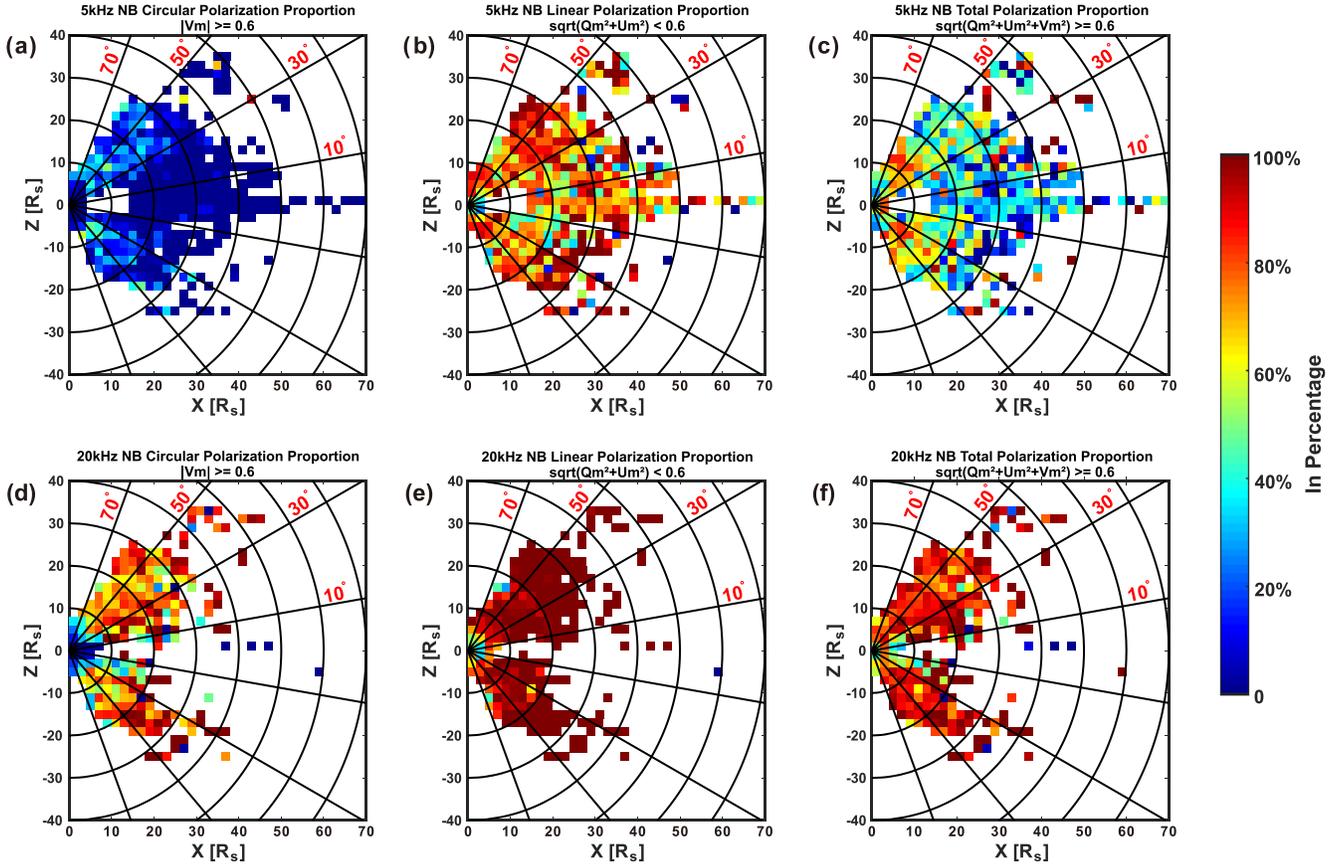

**Figure 4.** Panels (a)–(c) give the proportion of circular polarization, linear polarization, and total polarization for 5 kHz NB emissions with the absolute value of Stokes V larger than 0.6, linear polarization degree below 0.6, and total polarization larger than 0.6. Panels (d)–(f) give the results for the circular polarization, linear polarization, and total polarization of 20 kHz NB emissions in the same way as in panels (a)–(c).

polarization parameters directly, we give the proportion of the NB emissions with circular polarization >0.6, linear polarization <0.6, and total polarization >0.6 with respect to the total NB emission points in the corresponding bin. Figure 4(a) shows the proportion of circular polarization with absolute value larger than 0.6, while panel (b) gives the proportion of total linear polarization with value smaller than 0.6, panel (c) gives the proportion of total polarization with value lager than 0.6. Panels (d)–(f) give the results for 20 kHz NB emissions in similar format. We also studied the equatorial distribution for circular polarization, linear polarization, and total polarization characteristics, which are not shown here since no obvious difference like local time preferences resulted there. The equatorial distribution and other results can also be found in the Appendix (in Figures A4 and A6). Figure 4(a) shows that very few 5 kHz NB emissions have a high circular polarization with $|V| \geqslant 0.6$. At high latitudes and near the northern and southern hemisphere plasma torus edge, there are somewhat more highly circularly polarized 5 kHz NB emissions. However, the inversion algorithm will lose accuracy at small radial distances since the presumption of source location at the center of Saturn may be violated. So one should view the results within 10 $R_s$ with caution. The background noise around 5 kHz usually shows a high linear polarization degree compared to a normally small values for 5 kHz NB emissions as can be seen in Figure 1(c). This high linear polarization of the background can also pollute the 5 kHz NB emission signal. But after a careful check for each case identified by the program, we conclude that the 5 kHz NB emissions are either circularly polarized or unpolarized without linear polarization. Figure 4(c) shows that unpolarized 5 kHz NB emissions (or with a low total polarization) is very common in the equatorial plane and at low latitudes. One thing needs to be noted here which is that the polarization characteristics of NB emissions was extracted by using the goniopolarimetric data (e.g., data in Figures 1(b), (c)) inside the contour lines, which are obtained by using the algorithm described in Section 2. The contour lines are calculated based on the intensity spectra. The goniopolarimetric data mapped to the inside of the corresponding contour lines are not as clean as the intensity above 23 dB as shown in Figure 1(a).

One can easily recognize that the 20 kHz NB emissions are highly circularly polarized with many showing little linear polarization. It can be seen in Figure 4(d) that approximately 80% of NB emissions have circular polarization larger than 0.6, and in Figure 4(e) that roughly 95% of NB emissions have linear polarization smaller than 0.6. The 20 kHz NB emissions are highly circularly polarized, and the 5 kHz NB emissions are mostly unpolarized with a rather high proportion of circularly polarized events observed at high latitudes and at the northern and southern edges of the plasma torus. In this work, both 5 and 20 kHz NB emission events identified by the program also included the Z mode. According to other studies (Menietti et al.

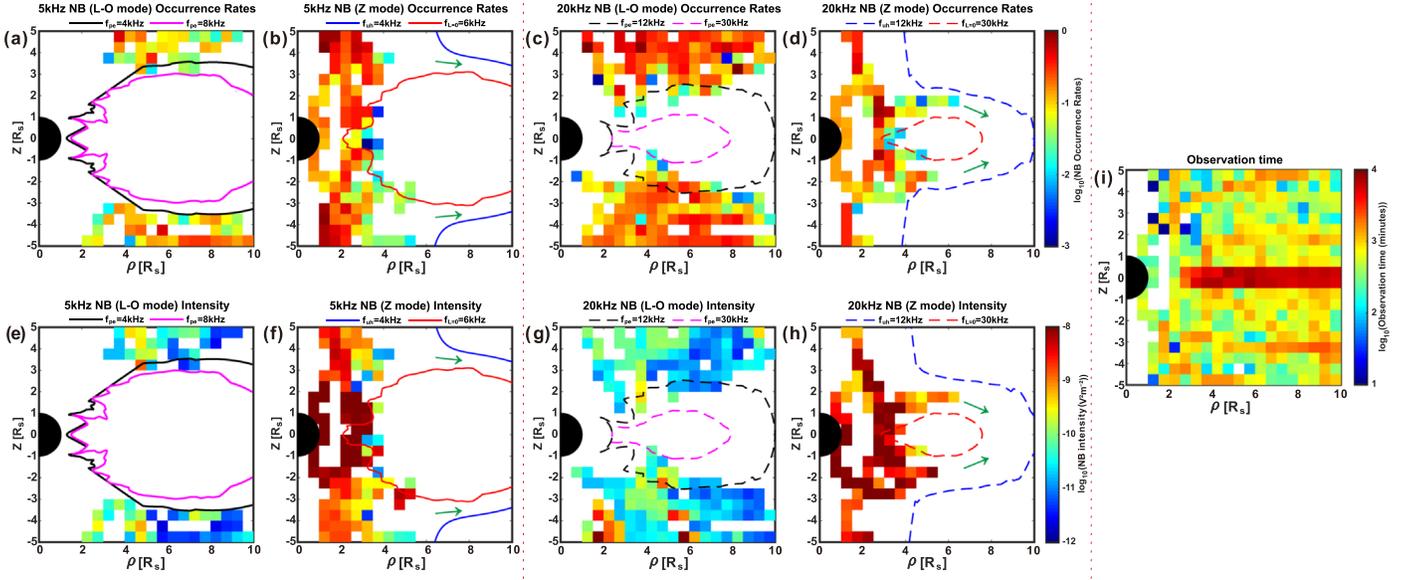

**Figure 5.** Panels (a)–(d) give the occurrence rates of the Z mode and L-O mode NB emissions distribution for 5 and 20 kHz, respectively, projected in the $\rho$–Z plane ($\rho$: distance from Saturn projected along the equatorial plane. Z: distance north/south of the equatorial plane, same as Figure 11 in Persoon et al. 2020). The bin size is in 0.5 $R_s$ × 0.5 $R_s$. Panels (e)–(h) give the intensity of the Z mode and L-O mode distribution projected into the $\rho$–Z plane. The black (solid in (a), (e) and dashed in (c), (g)) and pink (solid in (a), (e) and dashed in (c), (g)) contour lines are calculated based on a plasma torus density model, setting local fpe to 4 kHz (in (a), (e)) and 8 kHz (in (a), (e)) for 5 kHz NB emissions, and 12 and 30 kHz for 20 kHz NB emissions, respectively. The blue (solid in (b), (f) and dashed in (d), (h)) and red (solid in (b), (f) and dashed in (d), (h)) contour lines are the calculated $f_{uh}$ and $f_{L=0}$ frequency for Z mode NB emissions. The $f_{uh}$ lines in panels (b) and (f) are set to be 4 kHz for 5 kHz Z mode NB emissions and 12 kHz in panels (d) and (h) for 20 kHz NB emissions. The $f_{L=0}$ lines are 8 and 30 kHz for 5 and 20 kHz Z mode NB emissions, respectively. The calculation is based on the plasma torus density model of Persoon et al. (2020, 2006) and the magnetic field model of Connerney et al. (1982). The green arrows in panels (b), (d), (f), and (h) show the modeled channels that Z mode emissions can exist and propagate. Panel (i) gives the observation time of Cassini in this region.

2015, 2016), the 5 kHz Z mode waves are usually observed close to Saturn at regions with $f_{ce}$ larger than 5 kHz and at least 20 dB more intense compared to the L-O mode emissions (Ye et al. 2010b). The enhanced intensity regions near Saturn (within 6 $R_s$) in Figure 3 represent the Z mode waves in this study. For 20 kHz NB emissions, the Z mode was also found but with smaller occurrence rates close to Saturn (Menietti et al. 2015, 2018b).

In Figures 4(a) and (d) the positions with smaller circular polarization proportion close to Saturn within roughly 4 $R_s$ correspond to the similar region in Figure 3(a) and (c) with intensity enhancement. The Z mode should be responsible for these features and is discussed in more detail in the next section.

### 5. Discussion and Summary

According to the occurrence rates in Figures 2(a) and (c), we can see the cutoff of NB emission at the dense Enceladus plasma torus. The cutoff at the plasma frequency implies that the wave mode is L-O mode. We could estimate the density by calculating the plasma frequency at 5 and 20 kHz separately yielding densities of 0.31 and 4.96 cm$^{-3}$, respectively, using $f_{pe} = 8980\sqrt{n}$ with $n$ in cubic centimeters and $f_{pe}$ in hertz. These values are consistent with the statistical model given by Persoon et al. (2020) as shown in Figures 5(a), (c), (e), and (g).

The details of occurrence rates and intensity close to Saturn are shown by plotting the 5 and 20 kHz NB emissions with different wave modes separately. Based on the density model of Persoon et al. (2006, 2020) and the Zonal 3 magnetic field model of Connerney et al. (1982), we calculate the $f_{pe}$, $f_{uh}$, $f_{L=0}$ frequencies to study the spatial range of the L-O mode and Z mode. Since the L-O mode waves cannot propagate in

**Table 1**
The NB Emissions List

| Column | Units | Description |
|---|---|---|
| 1 | kHz | Frequency |
| 2 | … | Case number: the case number of each NB |
| 3 | year | S Year: Start year |
| 4 | day | S doy: Start day in the year |
| 5 | hour | S hour: Start hour |
| 6 | minute | S minute: Start minute |
| 7 | year | E Year: End year |
| 8 | day | E doy: End day in the year |
| 9 | hour | E hour: End hour |
| 10 | minute | E minute: End minute |
| 11 | … | Mode: wave mode of NB emissions, 0 for the L-O mode, 2 for the Z mode. |

(This table is available in its entirety in machine-readable form.)

areas with the plasma frequency higher than the waves' frequency, the L-O mode waves would be blocked at the boundary of the high density region as shown in Figures 5(a) and (c). The black and pink lines are the $f_{pe}$ from the density model. Panels (b) and (d) show that Z mode emissions are mainly distributed in the inner region of the plasma torus. The intensities of Z mode emissions in panels (f) and (h) are much higher than the L-O mode emissions. These results are consistent with the earlier findings (Ye et al. 2010b; Menietti et al. 2015). The contour lines in Figures 5(b) and (f) use the same parameters ($f_{uh}$ = 4 kHz and $f_{L=0}$ = 6 kHz) as Ye et al. (2010b). The channel formed by the isosurface of $f_{uh}$ and $f_{L=0}$ as shown in their Figure 9 is incompletely drawn in this work. The green arrows are used to indicate the channel.

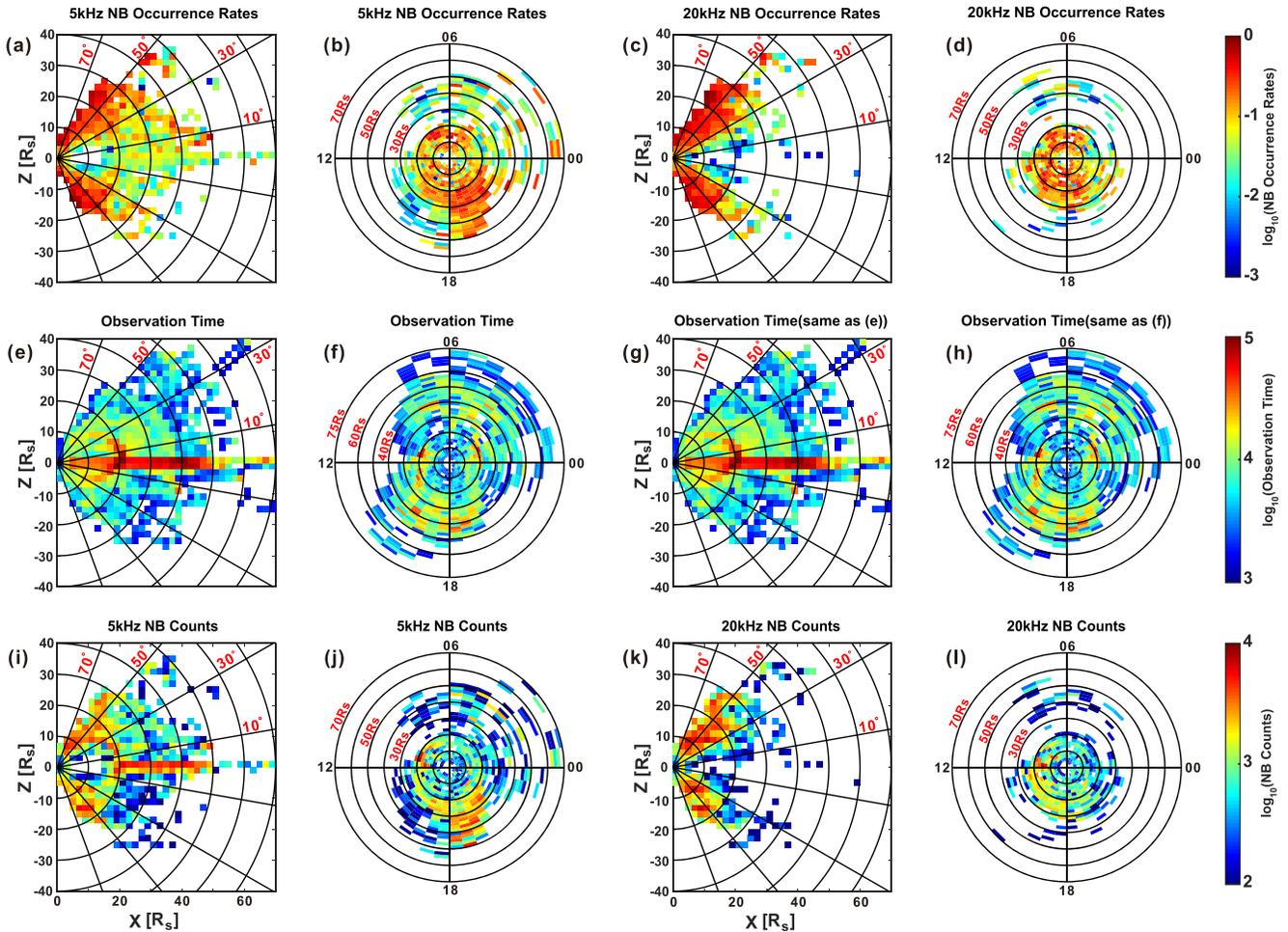

**Figure A1.** Panels (a)–(d) give the occurrence rates calculated by using the formula given in Section 3, and they are the same as panels (a)–(d) in Figure 2. Panels (e)–(h) show the observation time of Cassini RPWS from 2004, 001 to 2017, 255 in minutes projected into the meridional and equatorial planes. The color in each bin is in logarithm scale. Panels (i)–(l) give the valid NB emission points (in minutes) identified by the program. The occurrence rates in the upper row are obtained by dividing the number of points (minutes) in the lower row by the observation time in the middle row.

Interestingly, 20 kHz NB emissions are mostly observed at high latitudes, while 5 kHz NB emissions can propagate to low latitudes and in the region exterior to the plasma torus (i.e., roughly Lat < 10°, $R_s$ > 14$R_s$ in Figure 1(a)). The 20 kHz NB emission occurrence has been explained by Ye et al. (2009) that they would be reflected to high latitudes, due to the inner source location relative to the plasma torus. For 5 kHz NB emissions, as previously shown by Ye et al. (2010b) in their Figure 2 and also in Figure A2(a) in the Appendix, the intensity decreases with increasing radial distance in a $\frac{1}{R^2}$ dependence.

The 5 kHz NB emissions observed in the outer region of the plasma torus at low latitudes (e.g., at Lat < 10°, 14 − 20 $R_s$ in Figure 3(a) show intensities weaker than the 5 kHz NB emissions observed at high latitudes with larger radial distance (e.g., at Lat > 50°, 20 − 30 $R_s$ in Figure 3(a)). Therefore, the 5 kHz NB emissions in these equatorial latitudes in the outer region of the plasma torus do not seem to be a direct propagation result from the likely sources in the region of high intensity within 6 $R_s$. Additionally, Figure A3 in the Appendix shows that these low latitude NB emissions do not have a nice $1/R^2$ dependence as expected for direct propagation. It is currently unclear where the equatorial 5 kHz NB emissions beyond 15 $R_s$ comes from or from where it is reflected, and in the following we will suggest two different ideas.

The 5 kHz NB emissions within 6 $R_s$ show at least 20 dB intensity enhancement relative to the NB emissions observed at larger radial distance, which is due to the Z mode emissions in these regions. The first idea is that Z mode NB emissions could propagate outward along the edge of the plasma torus and mode convert to the L-O mode at the outer edge of the torus (Ye et al. 2010b), from where it can reach the equatorial regions at larger distance. The 5 kHz Z mode emissions usually occur in this region and have been suggested as important for the pitch-angle scattering and particle acceleration (Gu et al. 2013; Woodfield et al. 2018). Menietti et al. (2015, 2018a) gave a quite similar Z mode distribution within 8 $R_s$ compared to this study as shown in Figure 5(b). The electric and magnetic field intensities in their works all peak in the low latitude region. From the intensity distribution in the meridional plane in Figure 3(a), it seems that some 5 kHz NB emissions are produced close to Saturn and propagate toward the high latitudes. Another group of 5 kHz NB emissions, produced close to Saturn but through some different propagation process arrive at low latitudes outside the plasma torus while losing most of their intensity and polarization information.

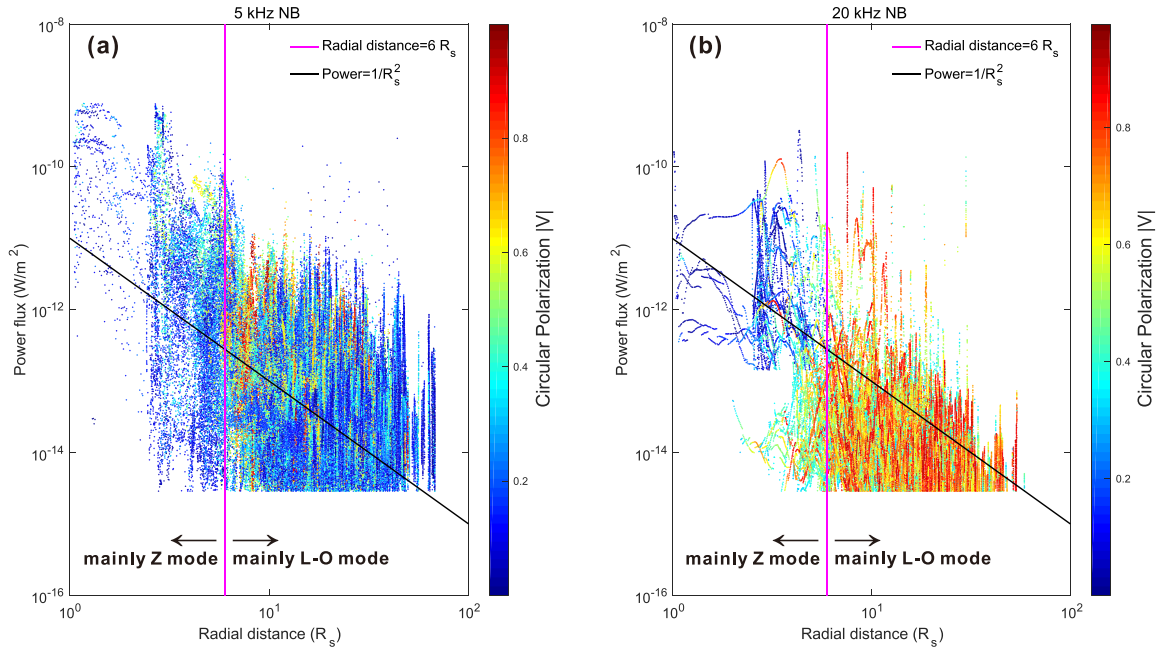

**Figure A2.** Panels (a) and (b) show the 5 and 20 kHz NB emissions in a similar format as Figure 2 of Ye et al. (2010b). The horizontal axis is radial distance in unit of Saturn radii. The vertical axis is the NB emission wave power. The black line is the $1/R^2$ dependence. The color represent the absolute value of the Stokes parameter (V). The pink line marks the 6 $R_s$ radial distance.

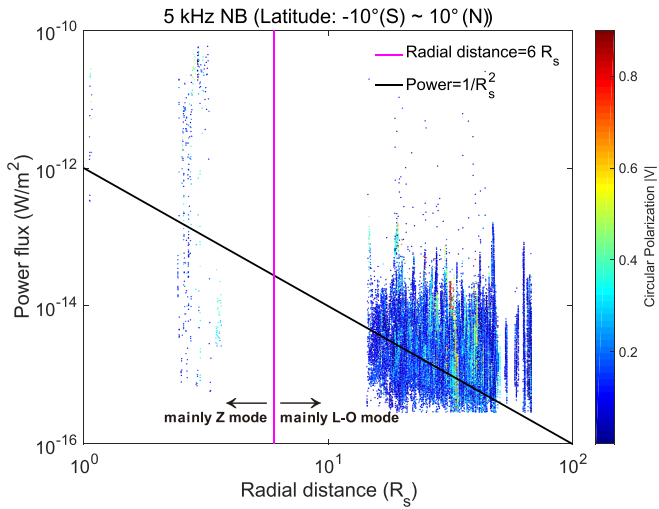

**Figure A3.** Low latitude 5 kHz NB emissions in a similar format as Figure 7. The 5 kHz NB emission events are confined within the low latitude regions ($-10°$ in the southern hemisphere to $10°$ in the northern hemisphere). The $1/R^2$ dependence line is the same as the black line in Figure A2.

As illustrated in Figure 4(a), the region at low latitude outside the plasma torus with a rather low proportion of circular polarization also shows weak intensity in Figure 3(a) and a low occurrence in Figure 2(a). Since the L-O mode 5 kHz NB emissions are mode converted from the Z mode at a density gradient and according to the model given by Ye et al. (2010b), the L-O mode 5 kHz NB emissions observed at low latitude might be mode converted from the Z mode emissions that traverse the narrow channel arriving at the outer edge of the channel outside the plasma torus. The Z mode is trapped between the isosurfaces of $f_{uh}$ and $f_{L=0}$ and it might propagate from one hemisphere to the other by reflections at those boundaries. Therefore, in order to propagate to the low latitude region outside the plasma torus the trapped Z mode would need to go through more reflections in the narrow channel to the outer edge of the plasma torus and thereby lose more energy and polarization. The 5 kHz NB emissions are mostly unpolarized at low latitudes, which may be explained by the reflections before the mode conversion from the Z mode to the L-O mode at a density gradient. The high latitude 5 kHz NB emissions would correspond to a mode conversion directly at the N-S edge of the plasma torus where the Z mode waves went through less reflection and a shorter propagation path before move conversion, and therefore the waves would show a higher intensity and circular polarization. The DF results of 5 kHz Z mode emissions show two possible source regions: one at high latitude field lines threading the SKR source region and the other at low latitudes inside the plasma torus (Menietti et al. 2018a). Therefore, these Z mode emissions excited at different positions may undergo different propagation paths.

The second idea is the reflection of the 5 kHz NB emissions at Saturn's magnetosheath. This seems possible as the thermal proton density in the magnetosheath shows a large variety from $0.01\,\mathrm{cm}^{-3}$ up to $2\,\mathrm{cm}^{-3}$ (Figure 7 of Sergis et al. 2013). The 25th and 75th percentile values for the density are 0.05 and $0.25\,\mathrm{cm}^{-3}$, respectively (Sergis et al. 2013). This means that, assuming quasi-neutrality, at least for some part of the time the electron density in the magnetosheath is high enough ($n > 0.3\,\mathrm{cm}^{-3}$) for a reflection of 5 kHz NB emissions, whereas it is always too low for a reflection of 20 kHz NB emissions (density $n$ of $2\,\mathrm{cm}^{-3}$ corresponds to 12 kHz). This could explain the low occurrence rate of less than 10% of equatorial 5 kHz NB emissions (Figure 2(a)), and maybe also the tendency of large distance 5 kHz NB emissions ($>50\,R_s$) to occur almost exclusively on the nightside (Figure 2(b)). And similar to the multiple reflections of the NB Z mode emissions, the reflection of the L-O mode at the magnetosheath could also lead to a decrease in the intensity and polarization degree. As the electron density in the magnetosheath changes gradually, there might be multiple refraction and reflection events before the wave coming from the Saturn-facing boundary of the plasma torus turns around and is radiated back toward Saturn's equatorial plane, thereby bypassing the plasma torus through which it cannot propagate.

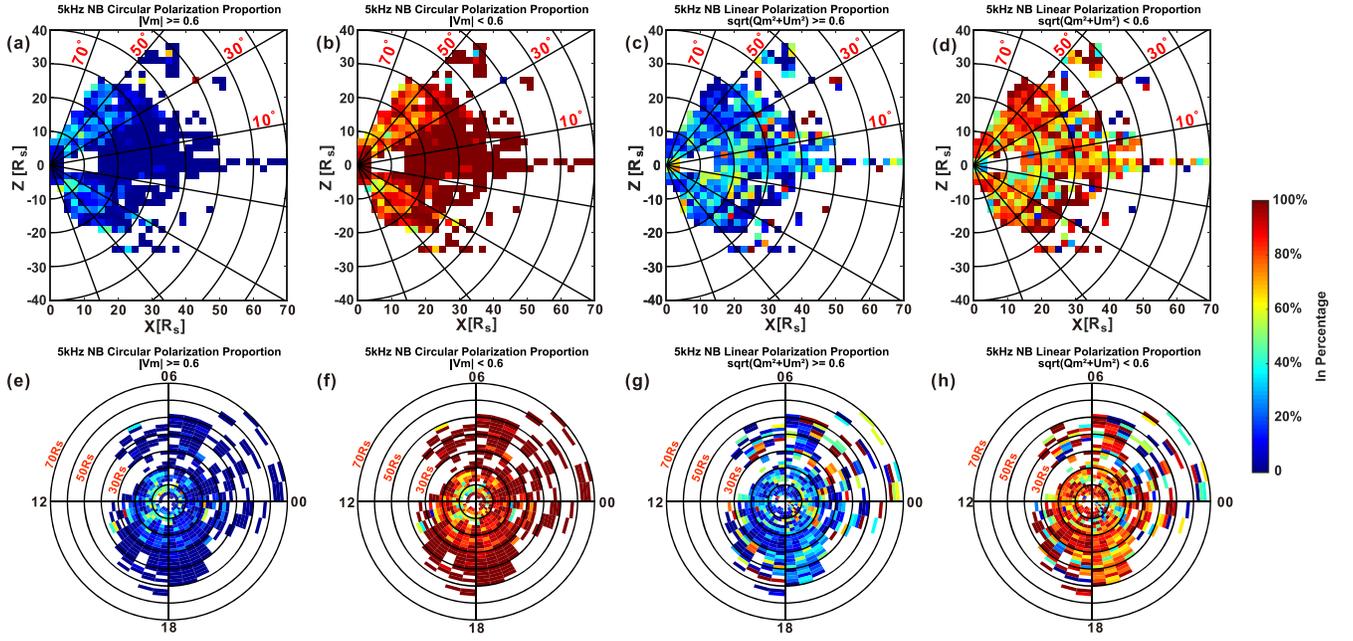

**Figure A4.** Panels, (a)–(d) give the circular and linear polarization proportion of 5 kHz NB emissions in the meridional plane. Panels (a) and (c) give the proportion of polarization parameters larger than 0.6 and panels (b) and (d) give the proportion of polarization parameters smaller than 0.6. Panels (e)–(h) give the proportion of circular and linear polarization of 5 kHz NB emissions in the equatorial plane.

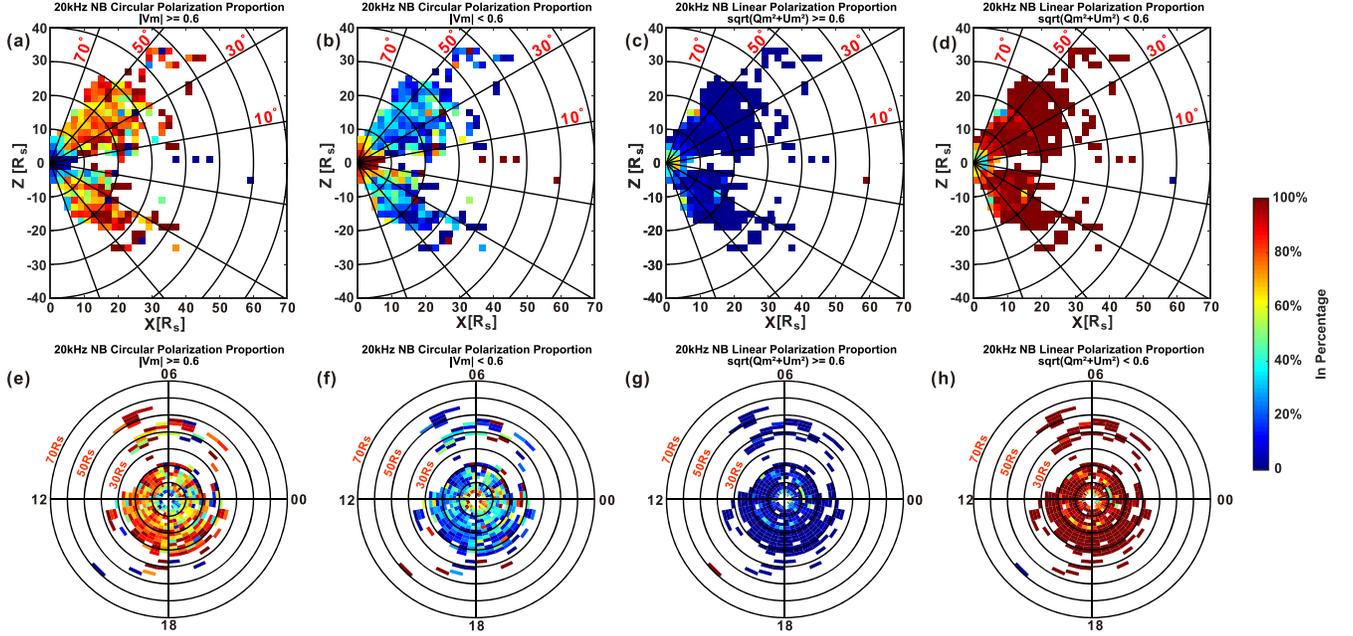

**Figure A5.** Similar format as Figure 9 but for 20 kHz NB emissions.

This mechanism can also explain the low circular polarization degree of 5 kHz NB emissions observed at high latitudes. The proportion of circularly polarized NB emissions observed at high latitudes shows that the 5 kHz NB emissions are less circularly polarized at high latitudes than the 20 kHz NB emissions. The reflection-by-sheath mechanism is consistent with the observation of the 5 kHz NB emissions at low latitudes outside the plasma torus during two magnetotail compressions (Figures 12 and 13 in Wang et al. 2010), during which the magnetosheath plasma density would be increased. The narrower bandwidth of 5 kHz NB emissions observed at low latitudes could also be explained by this mechanism because the highest frequency is limited by the plasma frequency in the magnetosheath. We checked the Cassini data and Voyager data. The Cassini data did not show any sign of 5 or 20 kHz NB emissions before the SOI (Saturn Orbit Insertion). The Voyager data did not show any detection of NB emissions before the Saturn flyby either (Gurnett et al. 1981). However, Figures 2 and 3 of this work indicate that there should be plenty of NB emissions observed beyond 30 $R_s$ on the dayside, outside the magnetopause. We will try to identify those events in future work and see if we could find any evidence of this new picture.

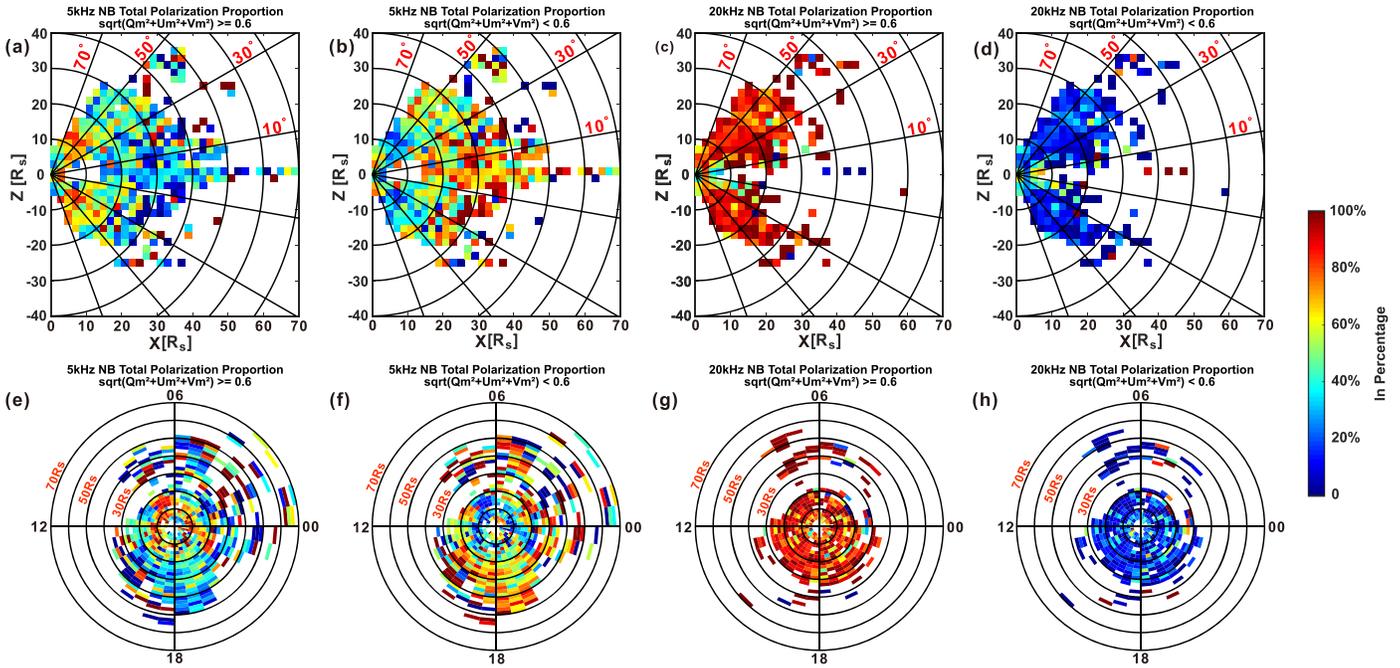

**Figure A6.** Panels (a), (b), (e), and (f) give the total polarization proportion of 5 kHz NB emissions in the meridional and equatorial planes. Panels (c), (d), (g), and (h) give the total polarization proportion of 20 kHz NB emissions in the meridional and equatorial planes. The colors represent the proportion percentage as described in Section 4.

As for 20 kHz NB emissions, very few low latitude cases were observed. However, as shown is Figure 5(d), there are some Z mode emissions extending to the channel formed by $f_{uh}$ and $f_{L=0}$. These Z mode emissions could propagate to the low latitude outside the plasma torus and mode convert to L-O mode emissions. According to the statistical results in this work, only 23 Z mode 20 kHz NB emissions were observed during almost 14 yr, so only a small number of Z mode emissions could traverse the channel. This may also imply that the 20 kHz NB emissions are mostly mode converted from the upper hybrid waves and near the boundary of the plasma torus. So, the emitted L-O mode would be reflected to higher latitudes and would not be affected by the magnetosheath as described above.

The 20 kHz L-O mode NB emissions converted from these Z mode emissions would also lose most of their intensity, polarization, and be observed at the low latitudes. Thus the 20 kHz NB emissions with a diffusive boundary, weak intensity, and low polarization degree would be hard for our algorithm to pick out. Both 5 and 20 kHz Z mode NB emissions close to Saturn tend to show a smaller proportion of the circular polarization and stronger intensity relative to the L-O mode NB emissions. The low circular polarization of Z mode emissions is due to the multiple reflections before mode conversion. The difference in polarization between the highly polarized 20 kHz L-O mode NB emissions and 20 kHz Z mode NB emissions also implies that these highly circularly polarized 20 kHz NB emissions may be directly mode converted from electrostatic upper hybrid waves.

Finally, the Z mode and O mode waves can also be generated directly in the SKR source region (Menietti et al. 2011) and from the temperature anisotropies near the inner edge of the plasma torus (Menietti et al. 2016, 2019). Then the generated Z mode and O mode emissions would go through the process described above.

Similar patterns of occurrence preference at the 18:00 ∼ 22:00 local time sector for 5 kHz NB emissions have been reported by Wang et al. (2010) This feature may be related to the nightside occurrence preference of the ENA injection events (Mitchell et al. 2015) and is not well understood. Wang et al. (2010) first suggested the link between ENA events (namely, the hot plasma cloud in their research) that were measured by the Magnetospheric Imaging Instrument (MIMI) Ion and Neutral Camera (INCA) (Krimigis et al. 2004) and the 5 kHz NB emissions, based on similar spatial locations, similar phase developments and onset times for these two phenomena. Recently Wing et al. (2020) studied the connection between the so called type 2 ENA injection event and the 5 kHz NB emissions and concluded that the type 2 injection can trigger the 5 kHz NB since the injection can introduce a plasma temperature anisotropy. The type 1 and type 2 injection events were categorized by Mitchell et al. (2015) on the basis of different physical processes (type 1: mainly related to reconnection at larger radial distance, type 2: mainly related to interchange instability at smaller radial distance), and they can be categorized easily according to the injection location. The relations between the injection events and the radio emissions in Saturn's magnetosphere are discussed in their work. Wing et al. (2020) explained that the 5 kHz NB emissions can be mode converted from upper hybrid waves at the outer boundary of the plasma torus and the temperature anisotropy can be provided by the injected hot plasma. According to the results in the present work, the intensity of both 5 and 20 kHz NB emissions peak on the Saturn-facing side of the plasma torus, which implies that the source location is close to Saturn. The interchange instability is originally caused by a the density gradient and an outward centrifugal force (Hill 1976; Siscoe & Summers 1981). The hot plasma carried by the flux tube would move inward through buoyancy force and would stop when the density gradient decreased to a low enough value. Lai et al. (2016) suggested that when an inward moving flux tube gets closer to the mass loading region ($L \approx 4$), the background cold plasma density increases much faster and compresses the flux tube immediately as the flux tube moves close to the plasma torus. To keep the pressure balance, more hot plasma carried by the flux tube would be squeezed to high latitudes along the magnetic field line. Therefore, this hot plasma being squeezed to high latitudes may introduce an electron temperature anisotropy and provide the free energy for the generation of NB emissions there. These injected hot plasma may also generate upper hybrid waves and then undergo the process described by Wing et al. (2020) at the same time. Wing et al. (2020) also showed the time lag between the plasma injection and the 5 kHz NB emissions. The time lag varied from a few minutes to 2 hr, which implied that the hot plasma may take different propagating paths to introduce the temperature anisotropies at the different possible source regions. These time lag differences are most likely caused by the different source locations (simply assume the propagation of the injected plasma started at the same location, then the time lag differences would come from the different propagation distances) and the time before mode conversion at a random location with density gradient. The detailed connection between NB emission and the plasma injection events will be investigated in future work. In the low latitude inner boundary of the plasma torus, a temperature anisotropy has been observed (Menietti et al. 2016; Wing et al. 2020) that can be responsible for the Z mode and O mode excitation, and the source is probably aided by ion cyclotron waves in this region interacting with the electrons.

By using almost 14 yr of Cassini RPWS data, a more comprehensive view of NB emissions has been obtained. The spatial distribution and polarization features of 5 and 20 kHz NB emissions were automatically detected by a computer algorithm by setting a series of criteria. The occurrence rates for both 5 and 20 kHz NB emissions show a high latitude preference and a 18:00 ∼ 22:00 local time preference for 5 kHz NB emissions. The 20 kHz NB emissions are predominantly observed at high latitude. The Z mode of both 5 and 20 kHz NB emissions' intensities peak within 6 $R_s$ and are with quite low polarization. Only 20% of 5 kHz NB emissions are highly circular polarized and these highly polarized emissions tend to be observed at the northern and southern edges of the plasma torus. The 20 kHz NB emissions are always highly circularly polarized with no linear polarization.

This work was supported by the Strategic Priority Research Program of the Chinese Academy of Sciences (grant No. XDB 41000000). The Cassini RPWS data used in this work was downloaded from the LESIA/Kronos collection with n2 level data (Cecconi et al. 2017a) and n3d data (Cecconi et al. 2017b). The magnetic field data from the Cassini MAG instrument was downloaded from the Planetary Data System at https://pds-ppi.igpp.ucla.edu/search/?sc=Cassini&t=Saturn&i=MAG.

Q8

Q9

## Appendix

The NB emissions list obtained by the computer program described in Section 2 is provided in machine-readable format for 5 and 20 kHz NB emissions. Table 1 shows the contents of the table including the frequency, the NB emissions case number in sequence of universal time (Case_number), the start time of the NB case (S_year, S_doy, S_hour, S_minute) and the end time of the NB emissions case (E_Year, E_doy, E_hour, E_minute), and the wave mode of the NB emissions in the last column (0: L-O mode, 2: Z mode). One can also find the obtained NB emissions list and contour lines data in this work for both 5 and 20 kHz NB emissions in Wu et al. (2021).

Figure A1 gives all the complement information of the occurrence of both 5 and 20 kHz NB emissions. As mentioned in Section 3, the total observation time and valid points of NB emissions are given. Figure A2 gives the NB emission intensities as a function of radial distance. To show the similar results obtained in this work and in Ye et al. (2010b). The NB emissions' intensities decrease as the radial distance increases and follow a $1/R^2$ power law. The red line marks a rough boundary of the Z mode NB emissions, which mainly distribute close to Saturn with radial distance smaller than 6 $R_s$ as also can be seen in Figure 5 in the text. Figure A3 gives parts of the data in Figure A2 in similar format, but only show the NB emission cases observed within 10° latitude. The results show that the low latitudes observed NB emissions with radial distance larger than 10 $R_s$ do not show a $1/R^2$ dependence as the radial distance increases. The results imply that these NB emission events' propagation paths are different from the high latitude events and may be due to the reflection at Saturn's magnetosheath as described in Section 5. Figure A4 gives the full information of the 5 kHz NB emissions with its circular polarization and linear polarization. Figure A5 gives the full information of the 20 kHz NB emissions with its circular polarization and linear polarization. Figure A6 gives the full information of the total polarization for 5 and 20 kHz NB emissions.

### ORCID iDs

Siyuan Wu ⓘ https://orcid.org/0000-0002-1326-8829
Shengyi Ye ⓘ https://orcid.org/0000-0002-3064-1082
G. Fischer ⓘ https://orcid.org/0000-0002-0431-2381
Jian Wang ⓘ https://orcid.org/0000-0002-7381-4713
Minyi Long ⓘ https://orcid.org/0000-0001-6984-5640
J. D. Menietti ⓘ https://orcid.org/0000-0001-6737-251X
B. Cecconi ⓘ https://orcid.org/0000-0001-7915-5571
W. S. Kurth ⓘ https://orcid.org/0000-0002-5471-6202